\documentclass[twocolumn,tighten]{aastex62}

\usepackage{xspace}
\usepackage{enumitem}
\usepackage{graphicx}
\usepackage{color}
\usepackage{comment}

\newcommand{\lya}        {Ly$\alpha$\xspace}
\newcommand{\hi}         {\ion{H}{1}\xspace}
\newcommand{\ha}         {H$\alpha$\xspace}
\newcommand{\oiii}       {[\ion{O}{3}]\xspace}

\newcommand{\unitcgssb}  {erg\,s$^{-1}$\,cm$^{-2}$\,arcsec$^{-2}$\xspace}

\newcommand{\unitcgslum} {erg\,s$^{-1}$\xspace}

\newcommand{\Llya}       {$L_{\rm Ly\alpha}$\xspace}
\newcommand{\llya}       {$L({\rm Ly\alpha})$\xspace}
\newcommand{\pol}        {\ifmmode{P_{\%}}\else$P_{\%}$\xspace\fi}

\newcommand{\kms}        {\ifmmode{\rm \,km\,s^{-1}}\else\,km\,s$^{-1}$\xspace\fi}

\newcommand\sig          {$\sigma$\xspace}
\newcommand\ff[1]        {\tablenotemark{#1}}

\definecolor{forestgreen}{rgb}{0.13, 0.55, 0.13}

\shorttitle{What Makes \lya Nebulae Glow?}
\shortauthors{Kim et al.}

\begin{document}

\title{
What Makes \lya Nebulae Glow? Mapping the Polarization of LAB\lowercase{d}05
}

\author{Eunchong Kim}
\affiliation{Department of Physics and Astronomy, Seoul National University, 
             Gwanak-gu, Seoul 88226, Korea}
\affiliation{Korea Astronomy and Space Science Institute,
             776 Daedeokdae-ro, Yuseong-gu, Daejeon 34055, Korea}

\author[0000-0003-3078-2763]{Yujin Yang}
\affiliation{Korea Astronomy and Space Science Institute,
             776 Daedeokdae-ro, Yuseong-gu, Daejeon 34055, Korea}

\author[0000-0001-6047-8469]{Ann Zabludoff}
\affiliation{Steward Observatory, University of Arizona,
             933 North Cherry Avenue, Tucson AZ 85721}

\author{Paul Smith}
\affiliation{Steward Observatory, University of Arizona,
             933 North Cherry Avenue, Tucson AZ 85721}

\author{Buell Jannuzi}
\affiliation{Steward Observatory, University of Arizona,
             933 North Cherry Avenue, Tucson AZ 85721}

\author{Myung Gyoon Lee}
\affiliation{Department of Physics and Astronomy, Seoul National University, 
             Gwanak-gu, Seoul 88226, Korea}

\author{Narae Hwang}
\affiliation{Korea Astronomy and Space Science Institute,
             776 Daedeokdae-ro, Yuseong-gu, Daejeon 34055, Korea}

\author{Byeong-Gon Park}
\affiliation{Korea Astronomy and Space Science Institute,
             776 Daedeokdae-ro, Yuseong-gu, Daejeon 34055, Korea}

%----------------------------------------------------------------------
\begin{abstract}
``\lya nebulae" are giant ($\sim$100\,kpc), glowing gas clouds in the distant universe.  The origin of their extended \lya emission remains a mystery. Some models posit that \lya emission is produced when the cloud is photoionized by UV emission from embedded or nearby sources, while others suggest that the \lya photons originate from an embedded galaxy or AGN and are then resonantly scattered by the cloud.  At least in the latter scenario, the observed \lya emission will be polarized. 
To test these possibilities, we are conducting imaging polarimetric observations of seven \lya nebulae. Here we present our results for LABd05, a cloud at $z$ = 2.656 with an obscured, embedded AGN to the northeast of the peak of \lya emission. 
We detect significant polarization. The highest polarization fractions $P$ are $\sim$10--20\% at $\sim$20--40 kpc southeast of the \lya peak, away from the AGN. The lowest $P$, including upper-limits, are $\sim$5\% and lie between the \lya peak and AGN. In other words, the polarization map is lopsided, with $P$ increasing from the \lya peak to the southeast. The measured polarization angles $\theta$ are oriented northeast, roughly perpendicular to the $P$ gradient.
This unique polarization pattern suggests that 1) the spatially-offset AGN is photoionizing nearby gas and 2) escaping \lya photons are scattered by the nebula at larger radii and into our sightline, producing tangentially-oriented, radially-increasing polarization away from the photoionized region. Finally  we conclude that the interplay between the gas density and ionization profiles produces the observed central peak in the  \lya emission. This also implies that the structure of LABd05 is more complex than assumed by current theoretical spherical or cylindrical models.

\end{abstract}


%----------------------------------------------------------------------
\section{Introduction}

\lya nebulae (aka ``\lya blobs'' or LABs) are rare, extended sources at $z$ = 2--6 with typical \lya sizes of 10\arcsec\ ($\sim$100 kpc) and line luminosities of \llya $\sim$ $10^{44}$ \unitcgslum \citep{Keel1999, Steidel2000, Francis2001, Dey2005, Yang2009, Yang2010}.  They lie in overdense regions of compact, \lya-emitting galaxies and generally have multiple, embedded sources \citep{Matsuda2004, Palunas2004, Yang2010, Prescott2008, Prescott2012, Badescu2017}. Comparison of an untargeted \lya survey with a large volume cosmological simulation \citep{Yang2009, Yang2010, Umehata2019} revealed that \lya blobs occupy halos that evolve into those of groups and clusters of galaxies today. If so, their embedded galaxies are likely merge to form brightest cluster galaxies at $z$ $\sim$ 0, and the \lya-emitting gas may represent the proto-intracluster medium. Determining the origin of the blobs' \lya emission is therefore essential to understanding the evolution of large-scale structure and the most massive galaxies.

%The \lya emission could arise through collisional excitation in a partially ionized medium. 
Some studies suggest that \lya blobs are produced by 
shocks from superwinds expelled by embedded AGN or starburst galaxies \citep{Taniguchi-Shioya2000, Mori2004,Geach2009, Cen2013, Cabot2016, Cai2017} or from cold gas accretion along filaments \citep{Haiman2000, Fardal2001, Nilsson2006, Goerdt2010, Faucher-Giguere2010, Rosdahl-Blaizot2012}. Yet observations \citep{Yang2011, Yang2014a, Yang2014b} detect only modest outflows ($<$200\,\kms) and only one possible instance of an inflow. Thus, it
is unlikely that 
strong shocks are the dominant mechanism for producing the \lya emission in \lya nebulae.

An alternate possibility is that photoionization by nearby or embedded sources leads to hydrogen recombination and \lya production throughout the nebula \citep{Haiman-Rees2001, Cantalupo2005}.  Yet a detailed investigation of eight blobs \citep{Yang2014b} reveals only two that contain a hard ionizing source capable of photoionizing the surrounding gas, a finding consistent with the low overall fraction of blobs with known AGN \cite[17\%;][]{Geach2009}{\footnote It is still possible that a hard-ionizing source may be obscured along the line of sight and nonetheless produce UV photons, perhaps capable of ionizing nebulae, along other directions. We also note that most targeted QSOs are surrounded by some extended \lya emission \citep{Borisova2016,Arrigoni-Battaia2019}.  }.
Another potential explanation is that \lya photons generated by galaxies or AGN are then resonantly scattered by the cloud \citep{Hayes2011,Beck2016}. Distinguishing between these two scenarios using only photometric and spectroscopic data is extremely challenging.

Mapping the {\it polarization} of the extended \lya emission provides a means of discriminating between photoionization and scattering. In the case where, 1) the entire nebula is photoionized, 2) recombination leads to the production of \lya photons at points throughout the cloud, and 3) there is no subsequent scattering, the observed \lya emission will not be significantly polarized. If, on the other hand, the photoionization region is relatively small, i.e., within or very near a galaxy, some of the escaping \lya photons will be scattered by the rest of the nebula and into our sightline.  This \lya emission will be polarized; the polarization strength will increase with projected distance from the source, and the polarization angles will be tangential to the direction of that gradient \citep{Lee-Ahn1998, Rybicki-Loeb1999, Dijkstra-Loeb2008, Trebitsch2016, Eide2018}.

These two scenarios represent the extremes; the reality might lie in between. For example, polarization could occur even in a highly-ionized region if it still contains enough neutral hydrogen gas over a large volume to produce a significant scattering probability.

There are only two previous studies that focus on mapping the polarization in \lya nebulae. Both suggest that scattering plays some role, but the details of the polarization patterns differ. Rings of highly polarized \lya emission (up to 20\%) are measured at 4\arcsec--8\arcsec\ ($\sim$ 45 kpc) from the center of SSA22-LAB1 \citep{Hayes2011}, a \lya blob at $z$ = 3.09, suggesting that the \lya photons are produced centrally and then scattered at large radii. From the previous study of \citet{You2017}, we detected comparably strong (up to 17\%) polarization out to $\sim$25\,kpc from the center in the \lya nebula surrounding the radio galaxy B3 J2330+3927 at $z$ = 3.087. Unlike in SSA22-LAB1, however, the significant polarization is observed only along the blob's major axis, along the radio jet. The polarization angles are aligned perpendicular to this direction. 

It is impossible to draw statistical conclusions about the origin of the \lya emission from polarization constraints in only two nebulae. To distinguish among powering models, more such observations are needed, particularly of nebulae with different potential source types (i.e., galaxies that include starbursts or AGN, that are radio-loud or -quiet) and locations relative to the peak of \lya emission. Therefore, we are conducting a polarization survey of seven \lya nebulae at $z$ = 2 -- 4 and with a range of likely powering sources and configurations. When combined with \lya photometric and spectroscopic data, and compared with state-of-the-art radiative transfer models (S. Chang et al., in prep.), these observations will point to the mechanism or mechanisms that illuminate \lya blobs.

In this paper, we present imaging polarimetry of the second target in our program, LABd05, a \lya nebula at $z$ = 2.656 \citep{Dey2005} with an embedded, obscured AGN. Unlike in SSA22-LAB1 and B3 J2330+3927, this AGN is spatially offset relative to the peak of \lya emission; a configuration with the potential to provide a new constraint on the complex radiative transfer between the AGN and the extended, surrounding gas. 
The AGN's energy output is capable of producing all the nebula's \lya photons via photoionization and recombination, suggesting that little or no polarization should be observed if photoionization is indeed the source of the \lya emission and the subsequent scattering is negligible.
In fact, an earlier measurement failed to detect significant polarization \citep{Prescott2011}. However, that constraint, $P = 2.6 \pm 2.8\%$, was within a single, large aperture of 8.2\arcsec\ diameter (65.6 kpc) and made with a small telescope (the Bok-2.3m) under poor seeing. Here we map LABd05 with the MMT-6.5m telescope\footnote{Observations reported here were obtained at the MMT Observatory, a joint facility of the University of Arizona and the Smithsonian Institution.} and the same, powerful SPOL imaging spectrometer used in our previous paper \citep{You2017}. 

This paper is organized as follows.  In Section \ref{sec:obs}, we describe our target, observations, data reduction, and polarization calculation.  In Section \ref{sec:results}, we present the polarization map of LABd05, including detections and upper-limits, and discuss the asymmetry and gradient of the polarization pattern. We also show the non-random orientation of the polarization angles. In Section \ref{sec:interpretation}, we suggest a  qualitative interpretation for these results. Section \ref{sec:conclusion} summarizes our conclusions. In the Appendix, we test how robust our measurements are against the uncertainties in the location of apertures and the image alignment. Throughout this paper, we adopt the cosmological parameters: $H_0$ = 70\,\kms\,Mpc$^{-1}$, $\Omega_{\rm M}=0.3$, and $\Omega_{\Lambda}=0.7$.

%----------------------------------------------------------------------
\begin{figure}
\epsscale{1.2}
\centering
\plotone{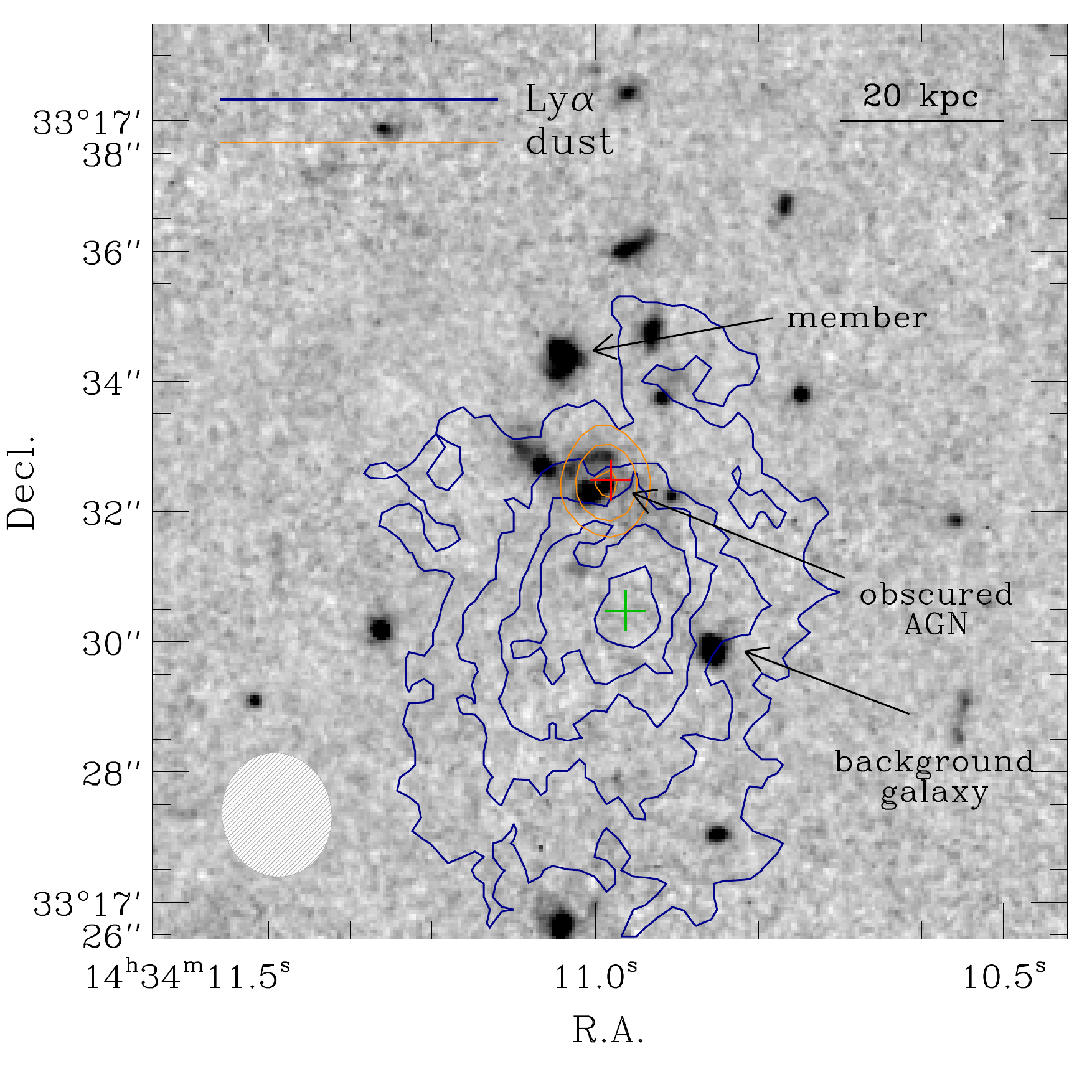}
\caption{{\sl HST VJH} composite image of LABd05 \cite[adopted from][Fig.~2]{Prescott2012} overlayed with \lya (blue) and 1.9\,mm dust continuum contours (red).  For clarity, we show only  the 5, 6, 7\sig contours of dust continuum emission.  \lya contours are plotted for 1, 3, 5, 7, 9\,$\times$\,$10^{-17}$ \unitcgssb. The hatched gray ellipse indicates the synthesized beam size for the dust continuum observation. The AGN (red cross) lies in a dusty region and to the northeast of the peak of \lya emission (green cross). At least one other galaxy, further to the northeast, is at the same redshift as the blob and AGN.
}
\label{fig:LABd05}
\end{figure}

%----------------------------------------------------------------------
\section{The Data}
\label{sec:obs}

%----------------------------------------------------------------------
\subsection{Target}

LABd05 is a giant \lya nebula at $z$ $\sim$ 2.656 that lies at R.A. = {14$^{\rm h}$}{34$^{\rm m}$}{10\fs963} and Dec. = $+$33\degr 17\arcmin 30\farcs48 (J2000). The extended \lya halo is very bright (\Llya = 1.7$\,\times\,10^{44}$ \unitcgslum) and spatially extended over $\sim$160\,kpc \citep{Dey2005}. Figure \ref{fig:LABd05} shows the {\sl HST} {\sl VJH} composite image of LABd05 illustrating its complex structure \citep{Prescott2011}, including contours of \lya surface brightness (blue) and dust ($\lambda_{\rm obs}$ = 1.9mm) continuum (red) \citep{Yang2014a}. 
Unlike the two previous \lya blobs with polarization maps, SSA22-LAB1 \citep{Hayes2011} and B3 J2330+3927 \citep{You2017}, LABd05 is associated with a bright mid-infrared galaxy at R.A. = {14$^{\rm h}$}{34$^{\rm m}$}{10\fs981} and Dec. = {33\degr}{17\arcmin}{32\farcs48} that hosts an obscured  radio-quiet AGN \citep{Dey2005,Prescott2011,Yang2014a} and is spatially offset (by $\sim$2\farcs0, 16\,kpc) from the peak of \lya emission.

By comparing the CO emission line profiles from the AGN and the spatially extended \lya and \ion{He}{2} $\lambda$1640 line profiles over the nebula, \citet{Yang2014a} measure a low outflowing gas velocity ($\sim$100\kms), thus excluding a model in which superwinds produce the \lya emission. The line ratio between CO($3-2$) and CO($5-4$) suggests that there is a large reservoir of low-density molecular gas \citep{Yang2014a}. LABd05 resides in a high density environment \citep{Prescott2008, Prescott2011}, suggesting that the embedded galaxies may evolve into the massive elliptical members of a galaxy group or cluster.  

%----------------------------------------------------------------------
\begin{deluxetable}{ccccc}
\tablewidth{0pt}
\tabletypesize{\small}
\tablecaption{LABd05 \lya Polarization Observations}
\tablehead{
%--------------------
\colhead{Date      }&
\colhead{Weather   }&
\colhead{Seeing    }&    
\colhead{Total Exp.}&
\colhead{Rot.
Ang.\ff{a}}\\[-1.5ex]
%--------------------
\colhead{(UT)    }&
\colhead{        }&
\colhead{(arcsec)}&
\colhead{(hour)  }&
\colhead{(degree)}
%--------------------
}
\startdata
2013--06--11  &  thin clouds  &  0.8 -- 1.3 &     3.5  &     180, 0 \\
2013--06--12  &  thin clouds  &  0.8 -- 1.3 &     3.5  &      0     \\
2013--06--13  &  thin clouds  &  0.8 -- 1.3 &     4.0  &      90    \\
2016--07--08  &  clear        &  0.8 -- 1.3 &     2.0  &      0        
\enddata
\label{tab:logs}
\tablenotetext{a}{Rotator angle of the SPOL instrument.}
\end{deluxetable}
%----------------------------------------------------------------------

%----------------------------------------------------------------------
\subsection{Observations}

To measure polarization properties of LABd05, we used MMT SPOL in imaging mode.  We refer readers to \citet{Schmidt1992} and \citet{You2017} for the details of this instrument and observing strategies.  We used the same filter and shim to achieve the central wavelength of 4446\,\AA\ as \citet{Prescott2011}.

Observations were carried out over four nights: UT June 11--13, 2013 and UT July 08, 2016. The weather was generally clear, although there was some cirrus during the run. The seeing ranged from 0.8\arcsec\ to 1.3\arcsec. We took 240\,s or 300\,s exposures for each position angle of the waveplate, thus each $Q$ and $U$ sequence was completed in 64\,m or 80\,m. Each $Q$ and $U$ sequence was taken at four waveplate position angles: $\alpha$, $\alpha$+90\degr, $\alpha$+180\degr, and $\alpha$+270\degr,  where $\alpha$ is the initial wave plate position angle, 0\degr\ and 22.5\degr, for the $Q$ and $U$ sequence, respectively. In total, we obtained 12 full sets of $Q$ and $U$ sequences  resulting in a total exposure time of 11.6 hours (excluding the bad quality images). The observation logs are in Table \ref{tab:logs}. 

%----------------------------------------------------------------------
\subsection{Data Reduction}

To reduce the data, we used our own {\tt IDL} reduction pipeline.  We first subtracted the overscan and performed flat-fielding using internal lamp flats and twilight flats. These flat images were taken with all the polarization optics (Wollaston prism and half-wave plate) in the optical path.  We used the {\tt L.A.COSMIC} package \citep{vanDokkum2001} to remove cosmic rays from our images. We examined the cosmic ray masks by eye to make sure that the real signal from the nebula remained.  During the run, we measured the polarization efficiency of the system ($p_{\rm eff}$ $\approx$ 0.973) by inserting a Nicol prism in the light path.  To place the observed linear polarization angle into an equatorial coordinate system, we obtained observations for the polarization standard stars BD+33 2642 and Hiltner 960.  We also observed the unpolarized standard stars G191B2B and BD+28 4211 to calibrate the narrowband fluxes and measure the instrumental polarization. The instrumental polarization was less than 0.1\%, which is consistent with our previous study \citep{You2017}.  

%----------------------------------------------------------------------
\begin{figure*}
\epsscale{1.2}
\centering
\plotone{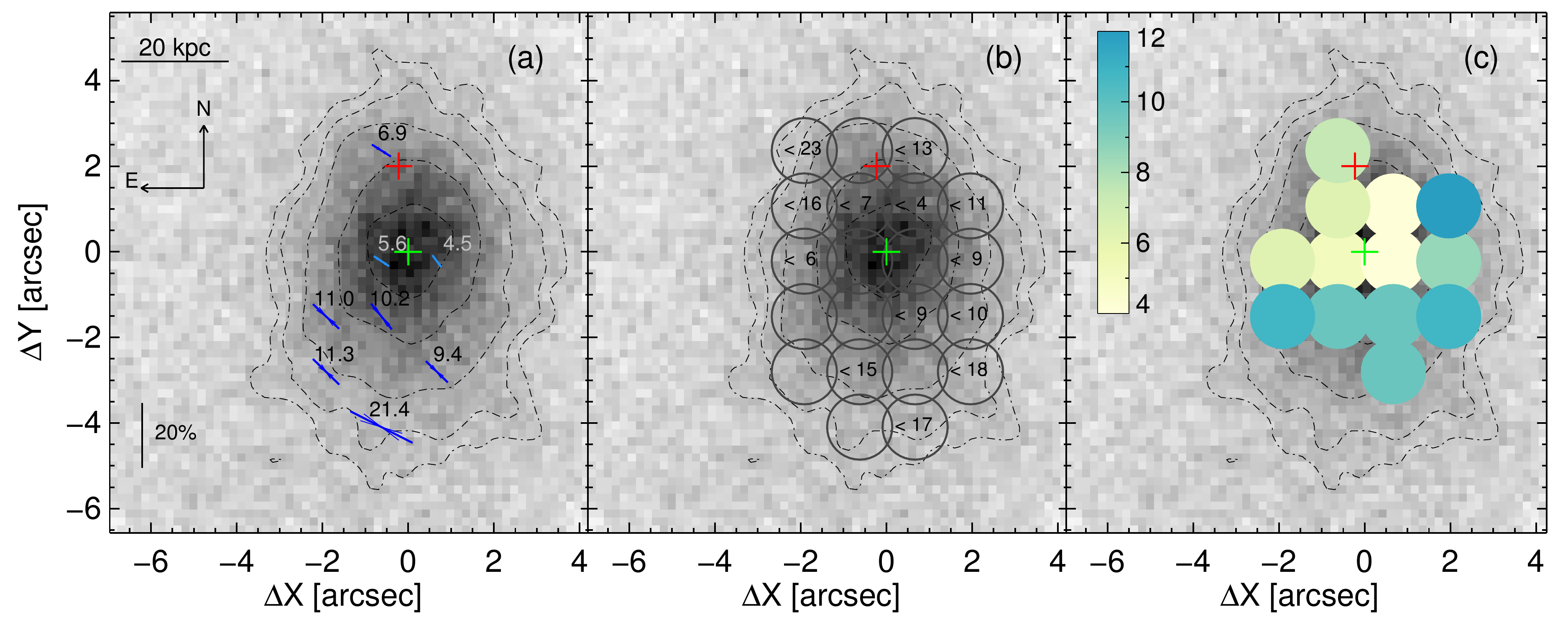}
\caption{Total (\lya plus continuum) polarization map overlaid on the \lya intensity map of LABd05. We show the \lya intensity (Stokes $I$) as a gray scale image in all three panels.  The dot-dashed contours represent the \lya surface brightness at 0.5, 1, 2, 4, and 6 $\times$ $10^{-17}$ \unitcgssb.  In each panel, the green cross represents the position of the peak of \lya surface brightness, and the red cross is the location of the obscured AGN, a potential energy source for powering the observed \lya emission.
{\bf (a)} The blue vectors are the eight significant ($\geq$2$\sigma$) polarization detections within the 1.52\arcsec\ (12\,kpc) sampling apertures; vector length and orientation indicate the degree of linear polarization $P$ and the angle of the polarization $\theta$, respectively. The $\pm1\sigma$ errors in $P$ and $\theta$ are overlaid as thin vectors around each of the main vectors.
The polarization pattern of LABd05 is asymmetric; most of the significant polarization is detected in the southeast of the nebula. There is gradient in $P$ for these detections that increases from the \lya peak to larger radii. The distribution of $\theta$ is not random (see also Figure \ref{fig:theta}); orientations to the northeast, perpendicular to the gradient in $P\%$, are favored.
{\bf (b)} The same map, but for the 2$\sigma$ upper limits. The gray circles represent the polarization measurement apertures used in all three panels. 
{\bf (c)} The same polarization map color-coded by the detected $P$ (panel a) and the upper limits (panel b) for only those apertures with $P \leq 12\%$. The weakest polarizations (yellow-green) lie mostly in the region between the \lya peak and the AGN.
\medskip
}
\label{fig:polmap}
\end{figure*}
%----------------------------------------------------------------------

%----------------------------------------------------------------------
\begin{figure}
\epsscale{1.1}
\centering
\plotone{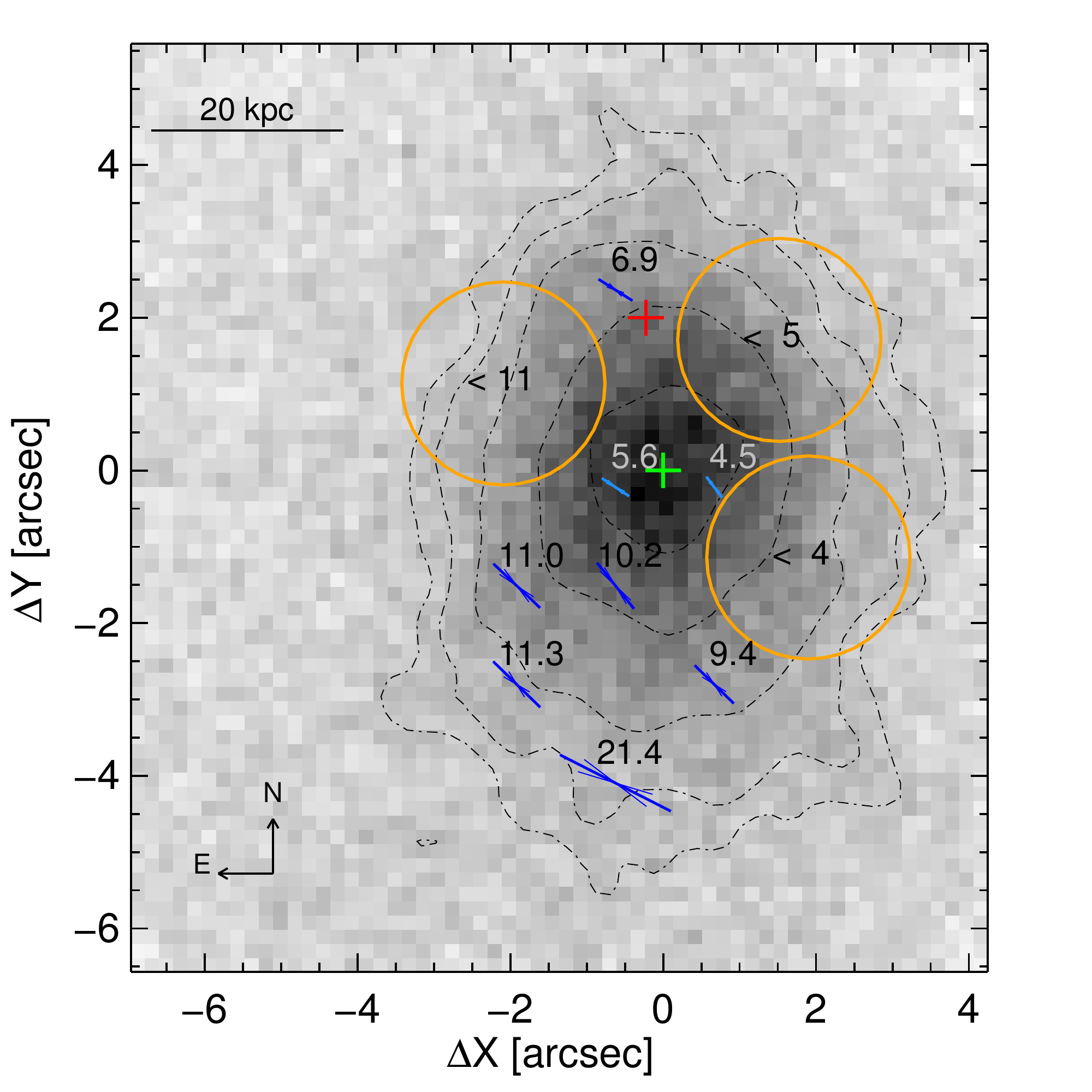}
\caption{
Example of three polarization measurements (orange) after applying a larger aperture size to where significant polarizations were not detected in Figure \ref{fig:polmap}. Although the aperture diameter here is expanded to $D$ = 2\farcs67 (21.3 kpc), there are still no significant polarization detections in those regions, only upper-limits.}
\label{fig:large}
\end{figure}
%----------------------------------------------------------------------

%----------------------------------------------------------------------
\subsection{Polarization Calculation}

We determine the polarization using the method in \citet{You2017}. Throughout the paper, $P$ and $\theta$ represent the fraction of polarization in percentage and the polarization angle on the sky measured from the North to the East direction, respectively. To minimize the spatial averaging of Stokes $Q$ and $U$ parameters and achieve the highest spatial resolution, we measure the polarization fractions $P$ within a grid of the smallest, fixed diameter (8 pixel, 1.52\arcsec, 12 kpc) apertures allowed by the seeing. Once we place the grid on the image, we consider only those apertures whose centers lie where there is \lya flux, in this case within the second outermost \lya surface brightness contour (1$\times$10$^{-17}$\unitcgssb) in Figure \ref{fig:polmap}.
In Appendix \ref{sec:aperture}, we describe how we test the sensitivity of the polarization map in Figure \ref{fig:polmap} to shifts in the location of this grid. 
In Appendix \ref{sec:image_combine}, we test how robust our polarization measurements are to shifts in image alignment between different, individual exposures. 

%----------------------------------------------------------------------
\begin{deluxetable}{cccccc}
\tablewidth{0pt}
\tabletypesize{\small}
\tablecaption{Polarization Measurements \\
}
\tablehead{
%--------------------
\colhead{ID  \ff{a}}&
\colhead{SB  \ff{b}}&
\colhead{$P$ \ff{c}       }&   
\colhead{$\sigma(P)$      }&
\colhead{$\theta$         }&
\colhead{$\sigma(\theta)$ }\\
%--------------------
\colhead{                 }&
\colhead{                 }& 
\colhead{($\%$)           }&
\colhead{($\%$)           }&
\colhead{(degree)         }&
\colhead{(degree)         }
%--------------------
}
\startdata
 1 &    12.9 &     5.6   &   2.1 &    57  &    10  \\
 2 &    12.6 &     4.5   &   2.1 &    37  &    13  \\
 5 &     9.5 &      10.2 &   2.8 &    39  &     8  \\
 8 &     6.9 &      11.0 &   3.9 &    47  &     10  \\
 9 &     6.7 &      6.9  &   3.4 &    58   &    14  \\
13 &    5.9 &      9.4  &   4.4 &    46  &    13  \\
17 &    4.3 &      11.3 &   5.6 &    46  &    14  \\
21 &    3.3 &      21.4 &   7.7 &    63  &    10 
\enddata
\label{tab:pol_list}
\tablenotetext{a}{Aperture IDs are defined in Figure \ref{fig:aperex}.  Each aperture has a diameter of $D = 1.52$\arcsec\ (12\,kpc).}
\tablenotetext{b}{Surface brightnesses in unit of $10^{-17}$\unitcgssb.}
\tablenotetext{c}{Corrected for statistical bias at low signal-to-noise ratio ($S/N$), following \citet{Wardle-Kronberg1974}.}
\end{deluxetable}
%----------------------------------------------------------------------

%----------------------------------------------------------------------
\section{Results}
\label{sec:results}

\subsection{Detection of Polarization}
\label{sec:pol_total}

Figure \ref{fig:polmap}a shows the total polarization map of LABd05, which includes both the \lya and continuum emission that entered the narrowband filter. We detect polarization in eight different apertures at $\geq$2$\sigma$ significance. Elsewhere, we achieve strong (2$\sigma$) upper limits (Figures \ref{fig:polmap}b and \ref{fig:polmap}c).  In Table \ref{tab:pol_list}, we list the polarization properties for the apertures with significant detections. Throughout this paper, we assume that the polarization signal here is driven by the polarization of \lya. In \citet{You2017}, we separated the contributions of the continuum and \lya to the total polarization of the nebula B3 J2330+3927, showing that the \lya polarization dominates.

\begin{figure*}
\epsscale{0.85}
\centering
\plotone{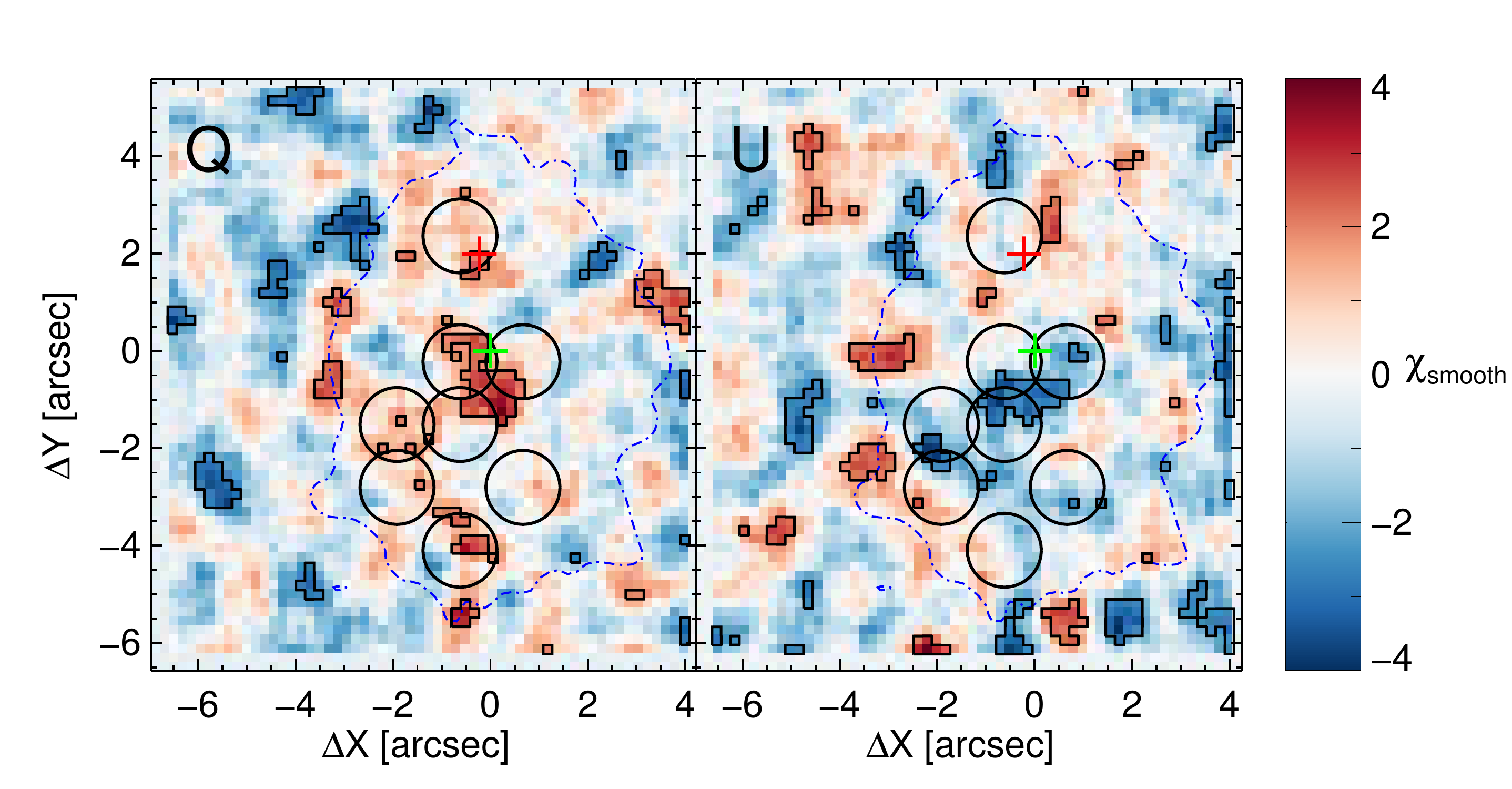}
\caption{$\chi_{\rm smooth}$ maps for the $Q$ and $U$ Stokes images.  Thick black contours outline the pixels with $|\chi_{\rm smooth}| > 2.5$, i.e., statistically significant regions. The dot-dashed contour is the outermost contour in Figure \ref{fig:polmap}. Black circles indicate the measurement apertures where significant ($\geq2\sigma$) polarizations are detected. The black circles are generally located around the outlined pixels, suggesting that these detections are real.
\medskip
}
\label{fig:stokes}
\end{figure*}

To test the significance of the detections in the eight apertures, we show the smoothed-$\chi$ image ($\chi_{\rm smooth}$) of Stokes parameter $Q$ and $U$ in Figure \ref{fig:stokes}. $\chi_{\rm smooth}$ is defined as in \citet{You2017}:
\begin{equation}
\chi_{\rm smooth} = \frac{I_{\rm smooth}}{\sigma_{\rm smooth}} 
                  = \frac{I_i \ast h(r)}{\sqrt{\sigma^2_{i} \ast h^2(r)}},
\end{equation}
where $I_{\rm smooth}$ is the image convolved with a tophat kernel $h(r)$ with a radius of 3 pixels. $\sigma^2_{\rm smooth}$ is the variance of the smoothed image propagated from the unsmoothed image. Given that $\chi_{\rm
smooth}$ should follow a normal distribution $N(0,1)$ for random noise,
$\chi_{\rm smooth}$ is useful to visualize low-$S/N$ features.
The eight apertures with significant detections in Figure \ref{fig:polmap}a are outlined as black circles here and generally coincide with the highest signal-to-noise pixels ($\chi_{\rm smooth}$ $>$ 2.5), suggesting that the detected polarizations are real.

In the following sections, we discuss the overall polarization pattern, including the asymmetry and radial gradient in the polarization fraction $P$, as well as the non-random distribution of polarization angles $\theta$. 

%----------------------------------------------------------------------
%----------------------------------------------------------------------

%----------------------------------------------------------------------
\subsection{Asymmetry of Polarization}
\label{sec:asym} 

The polarization pattern of LABd05 is asymmetric; in Figure \ref{fig:polmap}a, most of the eight $\geq$2$\sigma$ polarization detections are located southeast of the \lya peak. There is relatively weak polarization ($P \sim 5\%$), including the 2$\sigma$ upper-limits, in the region between the \lya peak and the obscured AGN (Figures \ref{fig:polmap}b and \ref{fig:polmap}c). This asymmetric polarization pattern is distinct from the two previous imaging polarimetric studies, which found rotational or axisymmetric distributions: the concentric pattern in SSA22-LAB1 \citep{Hayes2011} and the symmetry along the radio jet (and blob major-axis) in B3J2330+3927 \citep{You2017}.

 To further test for polarization in the regions of the nebula where there were no significant detections in Figure \ref{fig:polmap}, we remeasure the polarizations there within larger apertures, thereby producing higher $S/N$ (Figure \ref{fig:large}).
Even after expanding the aperture size to a diameter of D = 2.67\arcsec\ (21.3 kpc), we still do not detect significant polarization fraction to the northeast, northwest, or southwest of the \lya center.
%----------------------------------------------------------------------
\subsection{Polarization Gradient}
\label{sec:gradient} 

Among the significant detections in Figure \ref{fig:polmap}, $P$ increases to the southeast from $\sim$5\% (near the \lya peak) to $\sim$10\% (at $\sim$20 projected kpc) to $\sim$20\% (at the outermost aperture at $\sim$40 projected kpc). The spatial distribution of $2\sigma$ upper-limits is consistent with this radially-increasing, asymmetric gradient.

%----------------------------------------------------------------------
\subsection{Non-Random Polarization Angles}
\label{sec:histo}
\begin{figure}
\epsscale{1.2}
\centering
\plotone{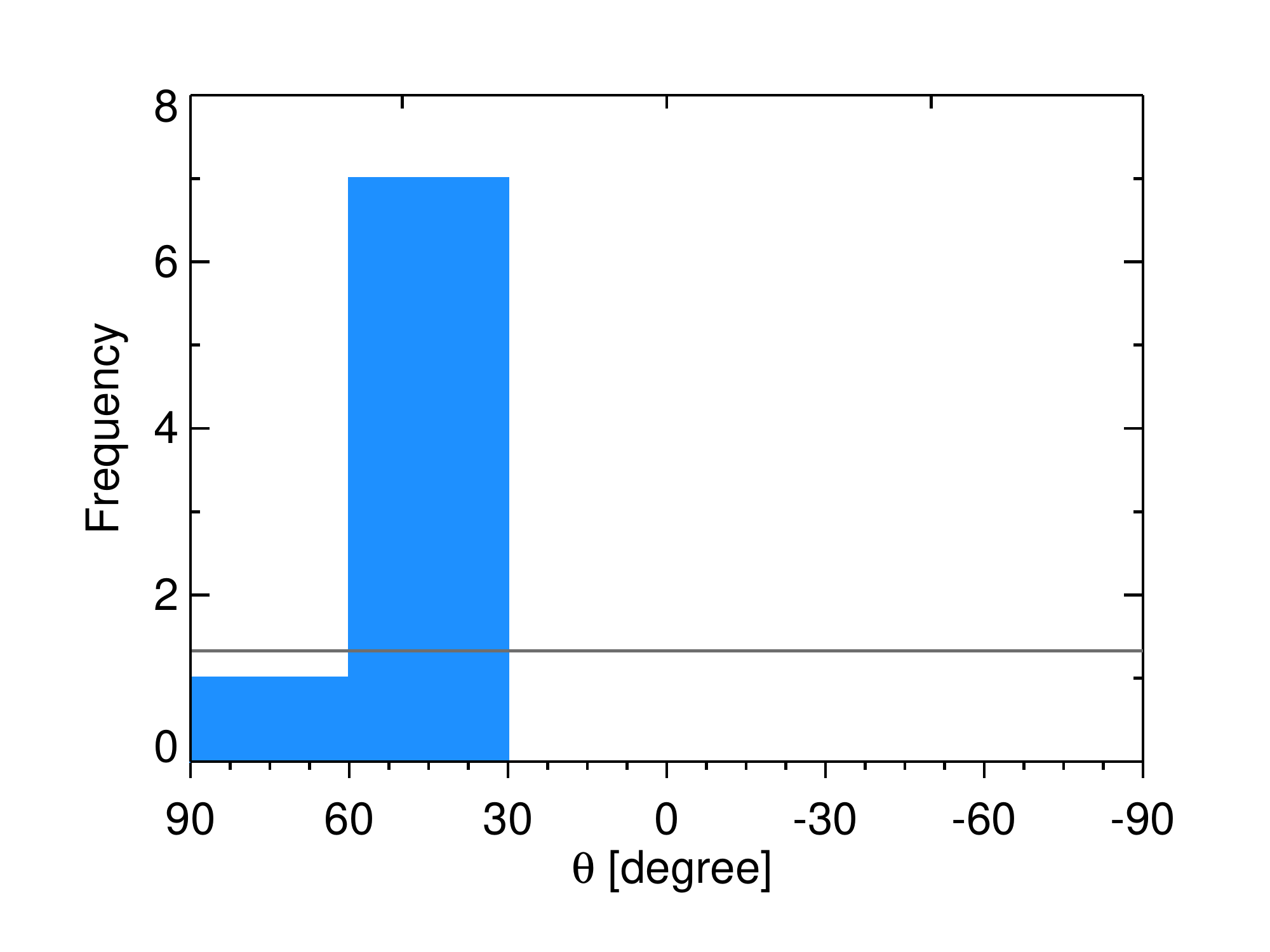}
\caption{Distribution of the polarization angle $\theta$ for the eight significant detections. The $\theta$'s tend to be oriented northeast (i.e., with a mean of +49$^{\circ}$), perpendicular to the direction of the polarization fraction gradient, and are distinguished at $>4\sigma$ from random alignments (grey horizontal line) by a Kolmogorov-Smirnov test. 
}
\label{fig:theta}
\end{figure}

The polarization angles $\theta$, defined from 0$^{\circ}$ (North) to +90$^{\circ}$ (East), are not randomly distributed in Figure \ref{fig:polmap}a; most appear to point to the northeast.  We test this result quantitatively in Figure \ref{fig:theta}. The distribution of $\theta$ is inconsistent (at $> 4\sigma$) with the uniform distribution expected at random and peaks to the northeast (i.e., close to +45$^{\circ}$).  The orientation of the polarization angles is then generally perpendicular to the gradient in $P$.

\subsection{Comparison to Previous Work}

\begin{figure*}
\epsscale{1.0}
\centering
\plottwo{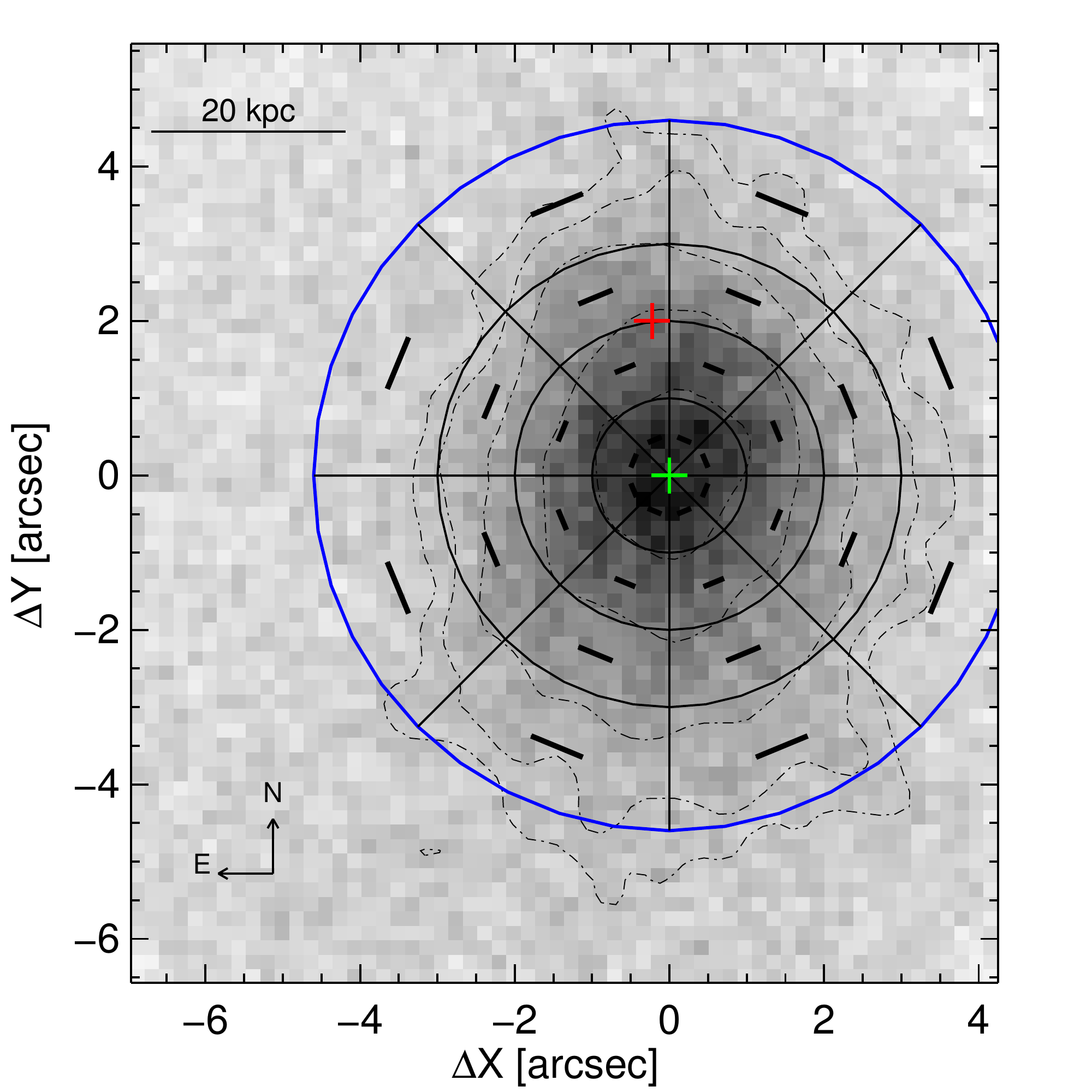}{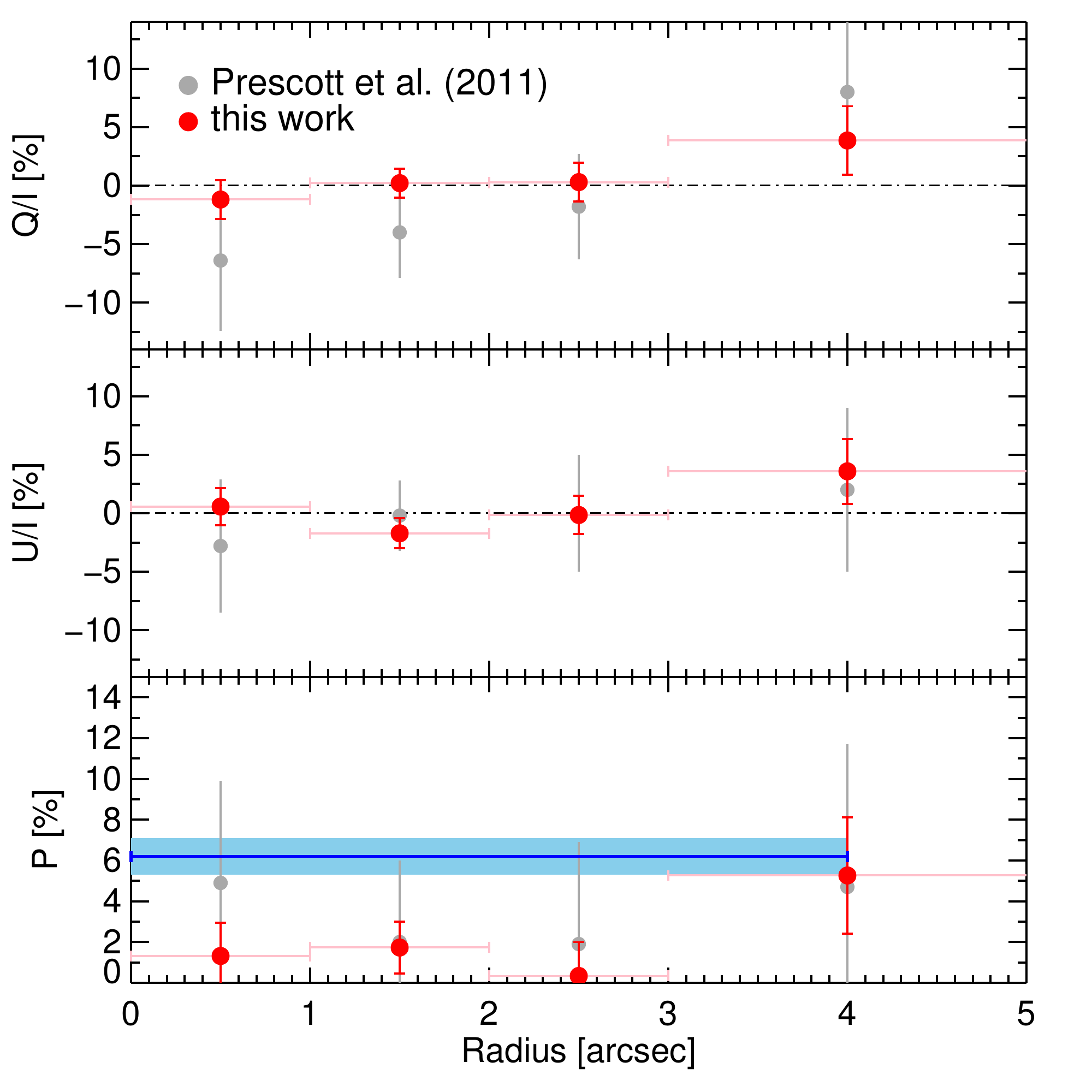}
\caption{Determination of radial profiles of the Stokes parameters and polarization fraction within each circular annulus about the \lya center.
{\bf Left:} \lya flux image from Figure \ref{fig:polmap} now overlaid with a schematic example of the polarization pattern (black vectors) predicted by some theoretical models \citep{Dijkstra-Loeb2008}. 
%First, we measure the polarization within a large-diameter (8\arcsec, 65\,kpc) circular aperture (blue outermost circle). 
%Second, at
At each radius, we measure the Stokes parameters ($Q$, $U$) within each azimuthal bin. Assuming that the polarization angles are as shown, we then
rotate those vectors to a fixed angle and average them to focus on the change in the strength of the polarization with radius.
{\bf Right:} Radial profiles of $Q/I$, $U/I$, and $P$ determined from our observations (red) and those of \citet{Prescott2011} (grey), who used the same azimuthal bins for their shallower measurements. The global polarization fraction within an 8\arcsec\ aperture (65\, kpc; blue outermost circle (left); blue shaded region (right)) is non-zero, and all three radial profiles ($Q$, $U$, $P$) are consistent with zero at all radii within the uncertainties.
\medskip
}
\label{fig:Moire}
\end{figure*}

To compare directly to the results of \citet{Prescott2011}, we measure two global polarization properties: the total polarization ($P_{tot}$) within a large 8\arcsec\ ($D$ = 65\,kpc) aperture (the same size used by \citealt{Prescott2011}) and the azimuthally-stacked radial profiles of the Stokes parameters.
 Our total polarization fraction is $P_{tot}$ = 6.2\%$\,\pm\,$0.9\% (Figure \ref{fig:Moire}), consistent with that in \citet[2.6\% $\pm$ 2.8\%]{Prescott2011}, which was formally a null detection. Our non-zero detection indicates a significant net polarization due to an asymmetry within the nebula.

 We determine the radial profiles
of the Stokes parameters
in same manner as \citet{Prescott2011}. We
measure polarizations within the azimuthal bins in Figure \ref{fig:Moire} (left), assuming that the polarization angles are as shown.
Before stacking to measure the change in polarization with radius, we align the polarization vectors within each annulus using the following coordinate transformation in the ($q$, $u$) plane:
\begin{eqnarray}
q_i^{\prime} &~=~& q_i \,\cos(2 \delta_{i}) - u_i \,\sin(2 \delta_{i}) \\
u_i^{\prime} &~=~& q_i \,\sin(2 \delta_{i}) + u_i \,\cos(2 \delta_{i}), \nonumber 
\label{eq:transform}
\end{eqnarray}
where ($q_i$, $u_i$) are the measured Stokes parameters in the $i$-th annulus and the $\delta_i$ is the position angle of the assumed concentric polarization vectors in Figure \ref{fig:Moire} (left).
Note that simple averaging over each annulus without alignment would wash out the polarizations even if there were a radially increasing pattern. But, this stacking method requires {\it a priori} knowledge of the orientation of the polarization vectors; here we assume a perfect concentric ring pattern for simplicity, but more detailed theoretical predictions should be used to test underlying physical models. 

Figure \ref{fig:Moire} (right) shows the resulting radial profiles of the normalized Stokes parameters $(Q/I)$\textsubscript{avg} and $(U/I)$\textsubscript{avg}, as well as the averaged polarization fraction within each annulus $P$. Although individual polarizations are detected across the nebula, all three radial profiles are flat and consistent with zero within the uncertainties, as they were in \citet{Prescott2011}. 

The above analysis demonstrates the importance of spatially-resolved imaging polarimetry in understanding the properties of \lya nebulae. Among the three \lya blobs studied to date, the measured global polarization fraction varies significantly, from a nearly unpolarized value of
1.9\%$\,\pm\,$0.9\% (within a 8\arcsec\ diameter aperture, 60\,kpc) in B3 J2330+3927
to $\sim$6\% (within 8\arcsec, 65\,kpc) in LABd05 
to $\sim$12\% (within 17\arcsec, 130\,kpc) in SSA22-LAB1.
Furthermore, low global polarization fraction does not preclude high local values; e.g., smaller areas within B3 J2330+3927 have up to $\sim$20\% polarization \citep{You2017}.
%%

%----------------------------------------------------------------------
%----------------------------------------------------------------------

%----------------------------------------------------------------------
%----------------------------------------------------------------------

%----------------------------------------------------------------------
\section{Discussion}
\label{sec:interpretation} 

%----------------------------------------------------------------------
\subsection{Current Models for \lya Nebulae}

Are the results of the previous section consistent with theoretical predictions? Models of \lya blob polarization generally assume that a neutral gas cloud surrounds a central \lya-emitting source (galaxy or AGN) from which radiation and winds flow or into which cloud gas falls \citep{Lee-Ahn1998, Dijkstra-Loeb2008, Eide2018}. \citet{Dijkstra-Loeb2008} investigate two simple scenarios: an expanding \hi shell and an optically thick, spherically symmetric, collapsing gas cloud. \citet{Eide2018} further explore the polarization of \lya emission for various geometries: static or expanding ellipsoids, biconical outflows, and the clumpy structure representative of a multiphase medium.
All of these models predict that the observed \lya polarization increases with projected radius in the nebula. This radial gradient arises because 1) photons at larger radii scatter by larger angles (i.e., closer to right angles) toward the observer, and 2) photons propagate more radially at larger radii due to the radiation field being increasingly anisotropic.

Symmetric polarization patterns are a natural consequence of these models. Concentric rings of  polarization are produced by either the expanding shell or spherical infall model of \citet{Dijkstra-Loeb2008}. Nebulae modelled with an ellipsoidal geometry and radial/bipolar outflows \citep{Eide2018} also predict a symmetric polarization distribution within the system. Unlike the mostly concentric and even axisymmetric polarization patterns observed in SSA22-LAB1 \citep{Hayes2011} and B3 J2330+3927 \citep{You2017}, respectively, LABd05's polarization morphology is asymmetric and a challenge for such models to explain.

%----------------------------------------------------------------------
\subsection{Explaining the Asymmetry}

The asymmetric polarization pattern in LABd05 suggests a more complex geometry than assumed by theoretical models of \lya nebulae to date \citep{Dijkstra-Loeb2008, Chang2017, Eide2018}. As discussed above, those models assume that the scattering \hi gas and \lya-emitting region are spherically or cylindrically symmetric, which generally produces symmetric polarization. Yet none of these models look like LABd05, which has no obvious source near its \lya peak and whose unique polarization pattern might arise from an offset between where the photons are generated and their scattering medium. 

The offset of LABd05's \lya peak from the AGN and the weak polarization region between them are consistent with this picture. If the AGN is photoionizing its immediate environs, as suggested by \citet{Yang2014a}, the gas between the AGN and the \lya peak could be highly ionized, allowing \lya photons escape without much scattering and with little polarization.  Some of these photons might then be scattered by neutral gas at larger radii, generating significant polarization far from the AGN. As discussed above, those photons that scatter into our sightline from the largest radii have the strongest polarizations. 
We would expect to observe increasing $P$ in the direction away from the photoionized region and $\theta$ perpendicular to that direction. The observed location of the peak of \lya emission---the blob's center---depends on the detailed structure of the photo-ionization region and the distribution of gas in the nebula.

 A caveat to this interpretation is that polarization could still occur in a highly-ionized region that retained some neutral hydrogen, because even small neutral fractions may produce a significant scattering probability over a large volume, due to the extremely large \lya cross section. We cannot rule out the possibility that the photo-ionized region in LABd05 might extend to larger distances, given the presence of the extended \ion{He}{2} and \ion{C}{4} emission lines. However, what neutral fraction or which ionization structure is required to produce any observable degree of polarization remains a mystery.

To fully understand the nature of LABd05 will require radiative transfer models with more realistic density and ionization profiles for the cloud and AGN (S.~Chang et al., in prep.). Ideally, such models will reproduce the photometric and spectroscopic observations as well, including the \lya surface brightness distribution on the sky, \lya emission line profile, and any velocity offset of the \lya line from non-resonant emission lines such as \ha and \oiii \cite[e.g.,][]{Yang2014b}.

%----------------------------------------------------------------------
\section{Conclusions}
\label{sec:conclusion}

We present imaging polarimetry of LABd05, a giant \lya nebula at $z$ = 2.656 with an embedded, radio-quiet, obscured AGN.  Our work here represents only the third such polarization mapping of a \lya nebula \citep{Hayes2011, You2017} and the first in which a possible powering source, the AGN, is spatially offset from the peak of \lya emission.

Our findings are:
\begin{enumerate}[leftmargin=0.5cm, itemsep=2pt]

\item 
We detect significant ($\geq 2\sigma$) polarization fractions $P \sim$5-20\% in eight different 1.52\arcsec\ (12\,kpc) apertures within the nebula.

\item
The polarization pattern is asymmetric; most of the significant polarization is to the southeast of the nebula.

\item
The weakest polarization ($\sim$5\%), including upper limits, is between the \lya peak and AGN. 

\item
$P$ increases outward from $\sim$5\% near the \lya peak to $\sim$20\% at $\sim$45 kpc projected to the southeast.

\item
The polarization angles $\theta$ are not randomly distributed and tend to point northeast, in a direction perpendicular to the gradient in $P$.

\item The total polarization fraction within an aperture of 8\arcsec\ (65\,kpc) diameter is non-zero,  $P_{tot}$ = 6.2\%$\,\pm\,$0.9\%, likely due to the blob's asymmetry preventing the local polarizations from cancelling out when added. Our value is consistent with that in \citet{Prescott2011}, which was formally a null detection due to its large uncertainties.

\end{enumerate}

The results above suggest a picture of LABd05 in which the gas between the AGN and \lya peak is highly photoionized by UV radiation from the AGN. \lya photons escaping this region along our sightline are not scattered and thus little polarized. Some \lya photons escaping in other directions are scattered by the neutral gas of the surrounding nebula at larger radii and into our sightline,  producing a polarization pattern that is generally tangential to and radially increasing from the \lya peak, with most of the significant polarization far from the AGN.

To date, only three high redshift \lya nebulae have been targeted for imaging polarimetric observations: SSA22-LAB1 \citep{Hayes2011}, B3 J2330+3927 \citep{You2017}, and LABd05 (this paper). These objects have a range of embedded potential powering sources and configurations, including multiple star-forming galaxies in SSA22-LAB1, a radio-loud, jetted AGN in B3 J2330+3927, and a radio-quiet, spatially-offset AGN in LABd05. 
Even this small sample yields key findings: 1) ubiquitous detection of significant polarization (up to $\sim$20\%), 2) tangentially-oriented, radially-increasing polarization gradients, and 3) a surprising diversity of polarization patterns ranging from concentric to axisymmetric to asymmetric, respectively.

The omnipresent polarization gradient, where the polarization is strongest at large radii, suggests that scattering plays a major role in \lya nebulae. On the other hand, the differences among the polarization patterns suggest variations in the gas density profile, velocity field, ionization structure, and/or location, energetics, and isotropy of the powering emission. Thus, future radiative transfer modelling should consider more complex geometries than spheroidal or cylindrical and strive to simultaneously predict the \lya surface brightness distribution, kinematics, and polarization pattern. On the observational front, an imaging polarimetric census of \lya blobs that span a range of potential powering source types and configurations is essential to explore and control for these effects. 

\smallskip
%-----------------
\acknowledgements
We thank the MMT staff and director, Grant Williams, for their assistance, and Arjun Dey and Moire Prescott for providing helpful information.
E.K. and Y.Y.'s research was supported by Basic Science Research Program through the National Research Foundation of Korea (NRF) funded by the Ministry of Science, ICT \& Future Planning (NRF-2016R1C1B2007782, NRF-2019R1A2C4069803).
A.I.Z. acknowledges funding from NSF grant AAG-1715609 and thanks the NRF and KASI for their support.
M.G.L. and E.K. were supported by the NRF grant (NRF-2019R1A2C2084019) funded by the Korea government (MSIT).
This work uses data that were obtained in part under the K-GMT Science Program (PID: 2016A-MMT-2) funded through Korean GMT Project operated by Korea Astronomy and Space Science Institute.

%----------------------------------------------------------------------
\smallskip
\facility{MMT (SPOL)}

%----------------------------------------------------------------------
\appendix

%----------------------------------------------------------------------
\section{Estimation of Uncertainties}
%----------------------------------------------------------------------
%----------------------------------------------------------------------
\begin{figure*}
\epsscale{1.0}
\centering
\plottwo{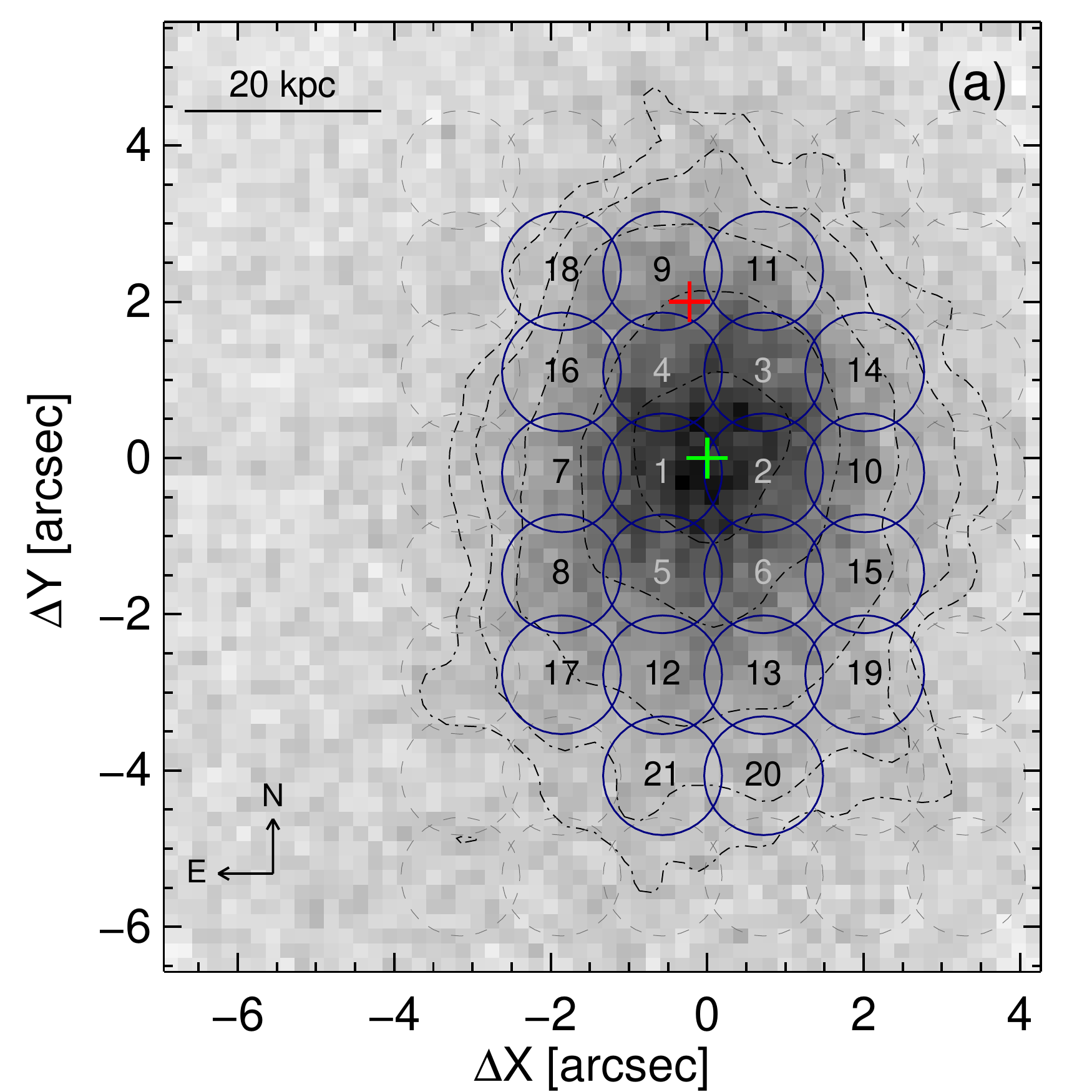}{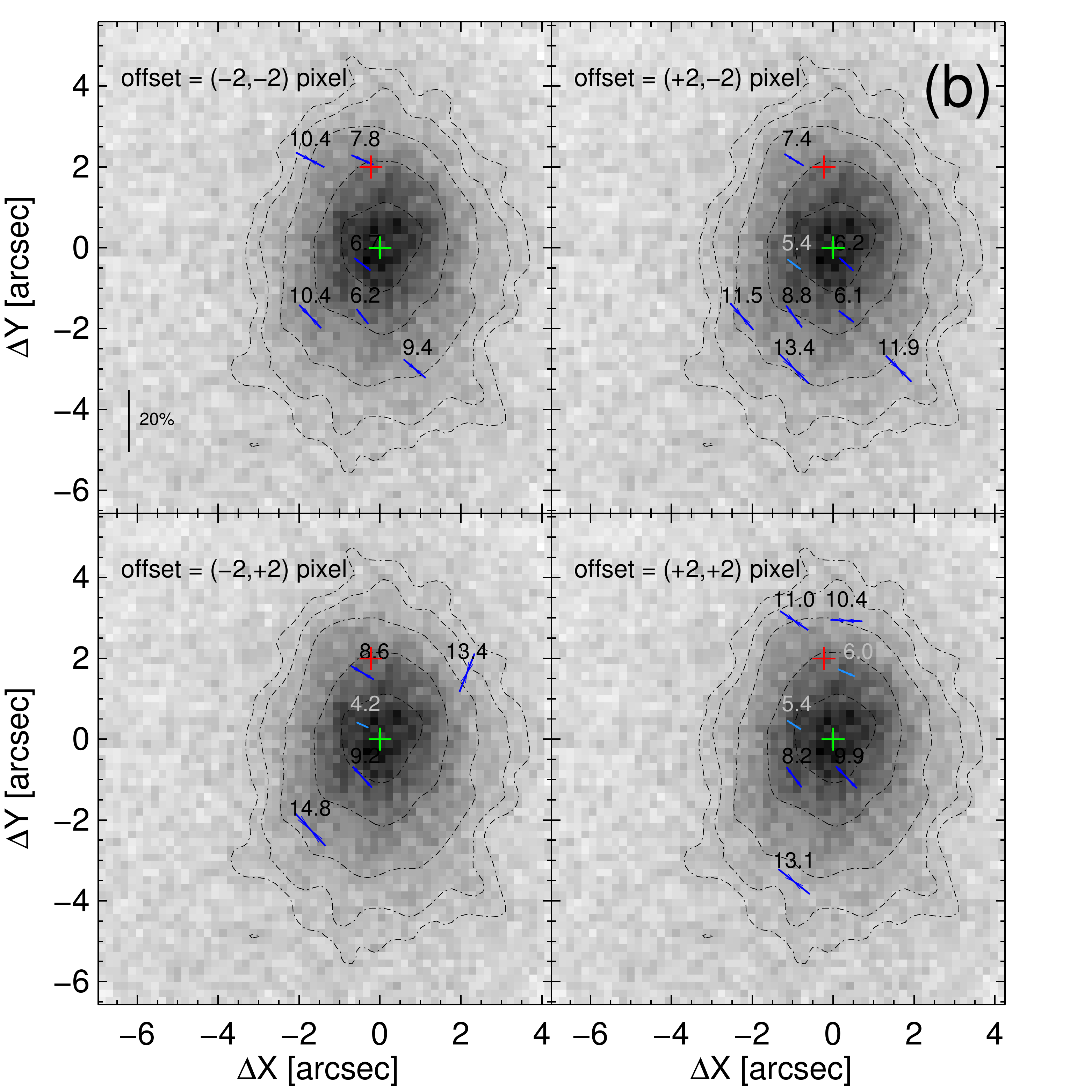}
\caption{
 Exploring the uncertainties arising from aperture placement.
{\bf (a)} Final grid of apertures (blue circles) adopted throughout the paper. The aperture IDs are assigned in order of \lya surface brightness; small ID apertures sample brighter parts of the nebula. To test how much the measurements are affected by the position of the grid, we shift the entire grid by ($\Delta X$, $\Delta Y$) pixels and remeasure the polarizations.
{\bf (b)} Four examples from this test. The overall polarization characteristics, such as the distribution of significant polarizations and the asymmetric polarization pattern, are similar in all four realizations and to Figure \ref{fig:polmap}, suggesting that our conclusions in this paper are robust against the location of measurement apertures.
}
\label{fig:aperex}
\end{figure*}
%----------------------------------------------------------------------

\subsection{Uncertainties due to Aperture Locations}
\label{sec:aperture}

Because the surface brightness of the nebula is low, we test how much the placement of the measurement apertures affects our results. First, we set up a grid of measurement apertures with the same diameter (8 pixel; 1.52\arcsec, 12 kpc) and a fixed spacing of 1.29\arcsec. This grid covers the entire nebula and adjacent sky background regions. All aperture centers lie within the second faintest \lya contour (1$\times$10$^{-17}$\unitcgssb) in Figure \ref{fig:aperex}a.
Then, we shift the entire grid by integer pixels ($\Delta X$, $\Delta Y$) in the $X$ and $Y$ directions, considering only those apertures whose centers lie within the second faintest \lya contour.

Figure \ref{fig:aperex}b shows four examples from this test: ($\Delta X$, $\Delta Y$) = ($-$2, $-$2), (2, $-$2), ($-$2, 2), and (2, 2). The significant polarization fractions in all four images are generally distributed to southeast part of the nebula, as in Figure \ref{fig:polmap}. Radially increasing polarization gradients are also seen in all four images. Therefore, we conclude that the placement of the measurement apertures does not significantly affect our results.

%----------------------------------------------------------------------
\subsection{Uncertainties due to Image Alignment}
\label{sec:image_combine}

%----------------------------------------------------------------------
\begin{figure*}
\epsscale{1.2}
\epsscale{1.0}
\centering
\plotone{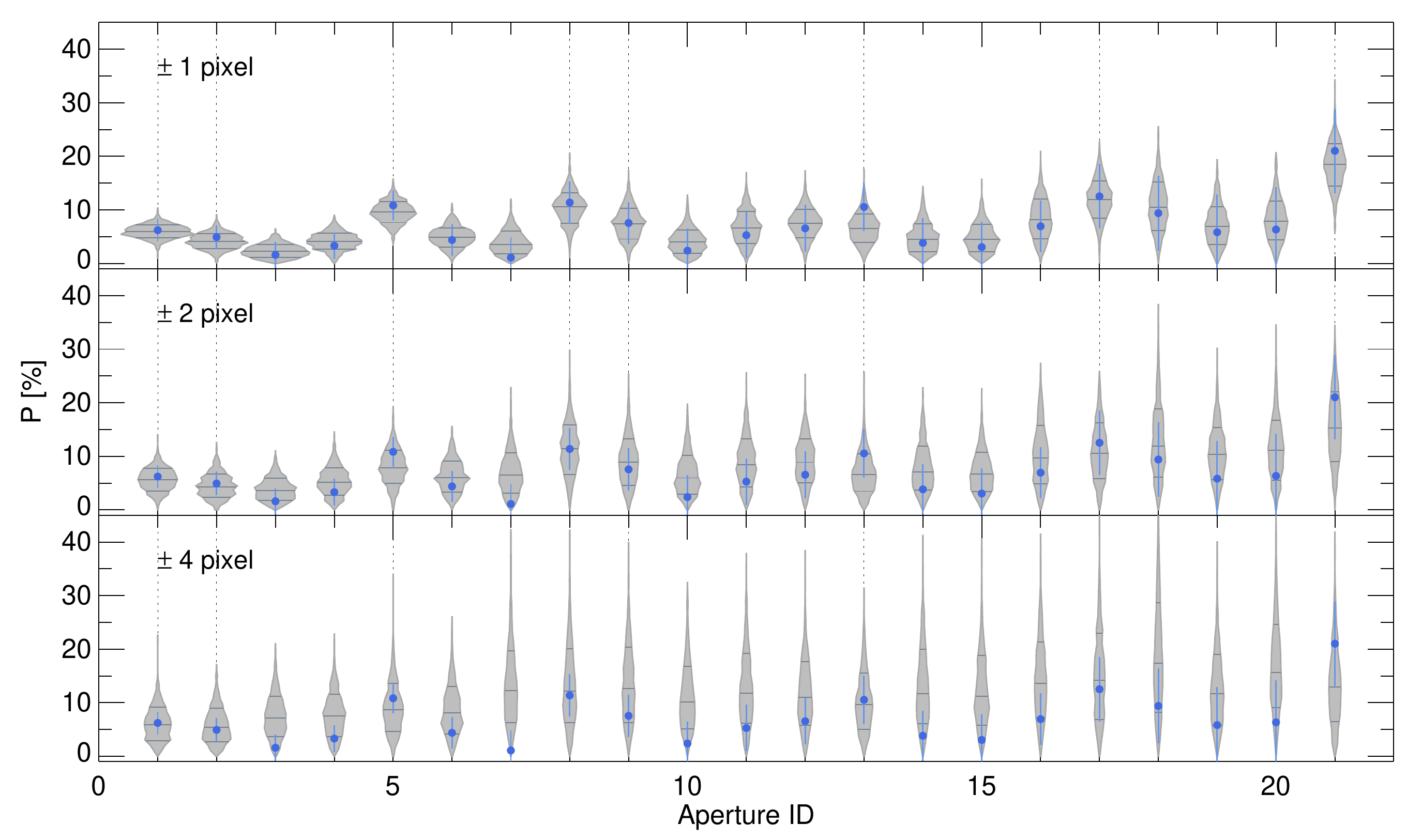}
\caption{ Assessing the uncertainties arising from image misalignment prior to combining the images. We show the distribution of polarization fraction $P$ for 1000 simulations with $\pm$1, $\pm$2, and $\pm$4 pixel misalignments from the best-aligned image (see Figure \ref{fig:polmap}). The three dark gray horizontal bars at each aperture ID represent the median and the 1$\sigma$ (68.3\%) range of the distribution. The blue dots with error bars indicate the polarization values for the best-aligned image.
The vertical dotted lines represent the apertures with significant ($\geq 2\sigma$) detections. In the case of small misalignments ($<$2 pixels; 0.4\arcsec), the variations due to the misalignment are smaller or comparable to the uncertainties associated with the best $P$ values (blue error bars), showing that small amounts of misalignment do not affect our results significantly.
\medskip 
}
\label{fig:alignment}
\end{figure*}
%----------------------------------------------------------------------

We align and combine the exposures using cross-correlation, because there is no other bright source that can be used as a reference in the SPOL field of view (19\arcsec\ $\times$ 19\arcsec). 
Here we check how much the image alignment procedure affects our results. We generate simulated sets of images by applying random $\pm$1 pixel shifts, reduce these misaligned data, and then measure polarization using the same method as for the original data. We repeat this process 1000 times, carrying out the same test with larger shifts of $\pm$2 and $\pm$4 pixels.

Figure \ref{fig:alignment} shows the results for $\pm1$, $\pm2$ and $\pm4$ pixel misalignments, respectively. As a function of aperture ID, we show the distribution of $P$ (shaded regions). The three gray horizontal bars for each aperture represent the median and the 1$\sigma$ (68.3\%) range of distribution. The vertical dotted lines represent the apertures with significant detections ($>$2$\sigma$) in the final best-aligned image. The variation of $P$ due to the {\it wrong} image alignment increases as the misalignment becomes larger.
In the case of a $\pm1$ or $\pm2$ pixel misalignment, the scatter is smaller or comparable to the measurement uncertainties associated with our final map (blue error bars). However, the scatter due to a misalignment of $\pm4$ pixels ($\pm$0.8\arcsec) becomes larger than these uncertainties, showing that the observed polarization pattern would be washed out. Therefore, we conclude that our results in this paper are robust against alignment errors within $\pm2$ pixels, which corresponds to $\pm$0.4\arcsec. Visual inspection of the alignment procedure shows that our alignment is typically better than $\sim$ 2 pixels.

%----------------------------------------------------------------------

\end{document}